# *Economics of Douglas fir management revisited*

Petri P. Kärenlampi*

Lehtoi Research, Finland
petri.karenlampi@professori.fi

* Author to whom correspondence should be addressed.

**Abstract**

A recent Douglas fir management investigation is repeated in terms of accounting measures. The rotation times become much shorter than in earlier results. Thinnings do not become feasible, provided the thinning effects on the volumetric yield function do not evolve in time. Evolving prices and expenses break the periodic boundary condition in monetary quantities: profit rates and capitalizations evolve with prices. The periodic boundary condition is, however, retained in time derivatives of dimensionless quantities, as well as in physical characteristics of rotations optimized for the rate of return. Then, optimal rotations do not depend on the evolution of prices and expenses. Relatively high timber prices shorten rotations, as relatively high expenses extend them.

## 1.Introduction

Forestry systems typically exhibit periodic patterns. In even-age forestry, the rotation age corresponds to the time difference between regeneration cuttings; in continuous-cover forestry, there is usually a rotation of harvest cycles [Rämö and Tahvonen 2017, Moog 2020]. Two different schools of thought exist in the economic discussion of rotation times: discounting of net revenues [Faustmann 1849, Fisher 1907, Fisher 1930, Hirschleifer 1958] or computation of some kind of return rate [Böhm-Bawerk 1889, Böhm-Bawerk 1851, Keynes 1936, Wright 1959]. The former is based on utility functions [Hirschleifer 1958], while the latter discusses wealth accumulation [Keynes 1936, Lutz and Lutz 1951]. The former appears more popular among forest economics, whereas the latter is financially better justified [Keynes 1936, Lutz and Lutz 1951, Boulding 1955, Chipman 1972, Chipman 1977, Hirschleifer 1958, Dorfman 1981]. It is also important whether computations are conducted on a cash-flow or accrual basis [Kärenlampi 2025a].

According to recent rigorous proof [Kärenlampi 2025a,b], suitable rotation times, thinning schedules, and regeneration, tending and fertilization practices do not depend on any discount rate. This finding changes presently common stand management considerations, based on the discounting of observables [Kilkki and Väisänen 1969, Haight and Monserud 1990, Pukkala et al. 2009, Tahvonen 2011, 2016, Rosa et al. 2018, Pukkala 2018, Tahvonen et al. 2010, Jin et al. 2019, Buongiorno et al. 2012, Tahvonen and Rautiainen 2017, Assmuth et al. 2021 Parkatti et al. 2019, Sinha et al. 2017, Rämö and Tahvonen 2015, Parkatti and Tahvonen 2020]. Correspondingly, a large body of forestry research must be redone, and novel results are difficult to compare with earlier literature.

The management practices of Douglas fir (*Pseudotsuga menziesii*) appear to be rather variable [Hudiburg et al. 2009, Nicolescu et al. 2023, Rossi and Kuusela 2023, Susaeta 2025, Kyaw et al. 2025]. The species is long-lived, and is able to form dense forests of large trees [Curtis 1995, Hudiburg et al. 2009, Nicolescu et al. 2023]. The volumetric growth rate declines relatively slowly along with age [Curtis 1995, Hudiburg et al. 2009, Nicolescu et al. 2023]. Suppressed trees do not necessarily recover well after thinnings from above [Emmingham et al. 2007]. The applied rotation periods are highly variable [Nicolescu et al. 2023].

In this paper, we discard compounding equations in the microeconomics and turn to accounting procedures [Speidel 1967, Speidel 1972, Kärenlampi 2025a]. A basic assumption is that the realm of forestry contains stands of a variety of development stages, and all stages are relevant observation points. Even within a single stand, any stage of development constitutes a relevant observation point. This leads to discussion of expected values of observables within any rotation [Kärenlampi 2025a, b].

The data used in this study originates from Susaeta [2025]. Here, the expected value of the rate of return on capital is first computed as a function of rotation time. This allows the determination of the financially optimal rotation, in the absence of thinnings. Then, the feasibility of thinnings is investigated. Thirdly, the effect of evolving prices and expenses on profit rate, capitalization and rate of return are clarified.

## 2. Materials

The data, originating from [Susaeta 2025], contains calibrated models for timber volume, monetary yield, and thinning response, as well as price and expense information. In particular, the timber volume density model is given as [Hudiburg et al. 2009, Rossi and Kuusela 2023, Susaeta 2025, Kyaw et al. 2025]

$$V(t) = a[1 - \exp(m\ t)]^c \qquad (1),$$

where $t$ is stand age, and other symbols refer to fitted parameters. Parametrization by Susaeta yields the timber volume density in thousands of board feet per acre [Susaeta 2025]. Development of timber volume density as a function of stand age is shown in Fig. 1.

The only price information not given in the paper by Susaeta [2025] is the market value of bare forest land. It is here taken the same as the unit expense of new stand establishment in the case of high productivity stands, and half of the establishment expense in the case of low productivity stands.

Interpretation of the thinning response model of Susaeta [2025] is more difficult. Firstly, thinning reduces the productive biomass. Secondly, increased availability of space increases the growth rate of remaining trees. Consequently, the thinning response model is hereby reformulated as

$$V_a(t+1) = V(t+1) * [1 - \frac{h(t)}{V(t)}] * [1 + \frac{h(t)}{V(t)} \delta] \qquad (2),$$

where $V_a(t+1)$ is the volume after thinning, and $h(t)$ is timber volume removed in thinning. The second factor on the right has been dimensionally corrected, and the third factor on the right has been made dependent on thinning intensity. Without such dependency, even an extremely gentle thinning would suffice in increasing the growth rate of the remaining trees by a constant factor $(1+\delta)$ [Kao and Brodie 1980, Susaeta 2025]. It is worth noting that the correction coefficients in Eq. (2) do not evolve in time [Susaeta 2025], unlike in the original postulation by Kao and Brodie [1980].

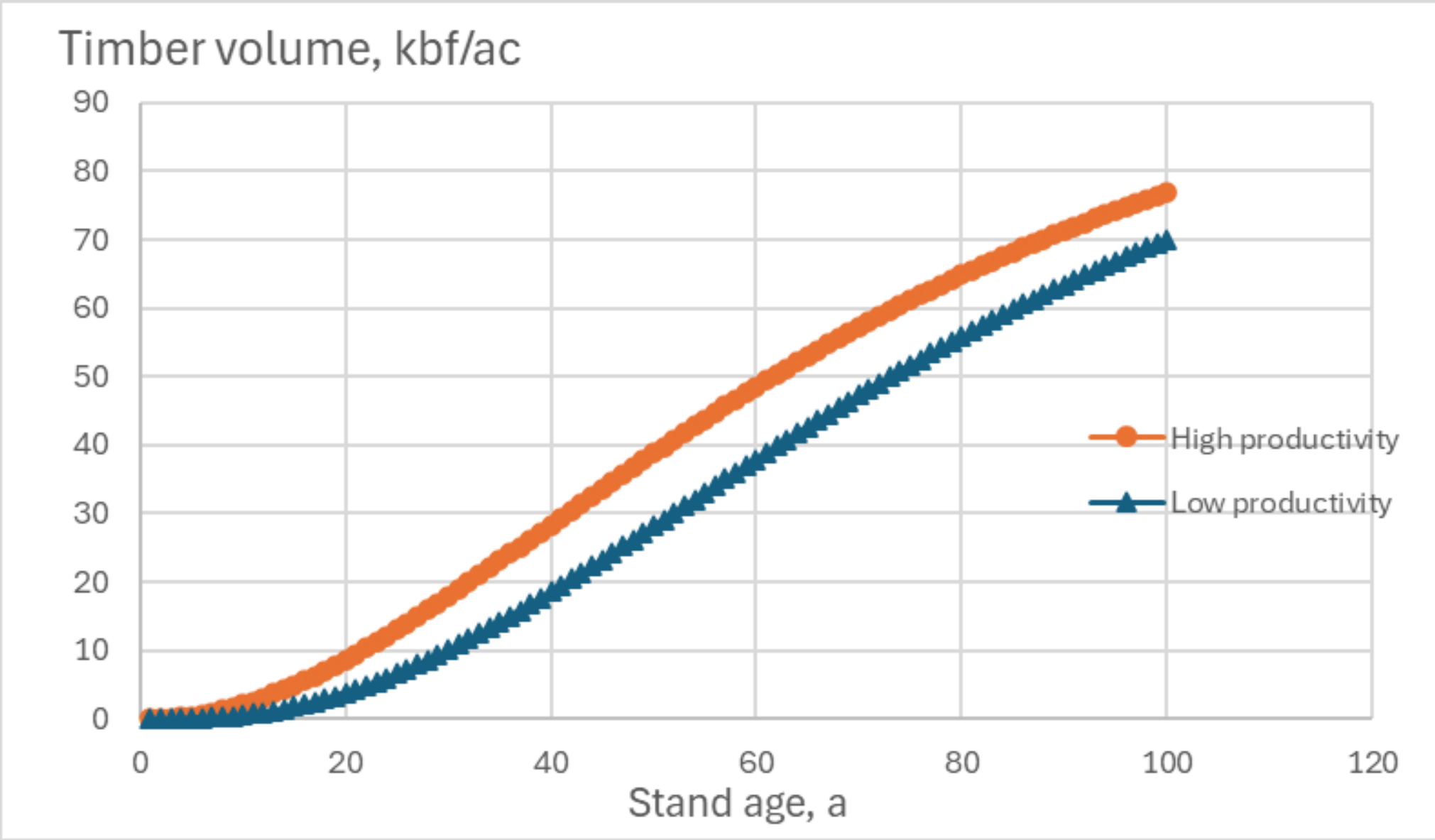

Figure 1. Timber volume in thousands of board feet per acre as a function stand age, according to model results of Susaeta [2025].

### 3. Computational methods

The return rate on capital is the relative time change rate of capital value. We choose to write

$$r(t) = \frac{d\kappa}{K(t)dt} \qquad (3)$$

where $\kappa$ in the numerator considers value growth, operative expenses, and amortizations, but neglects investments and withdrawals. In other words, it is the change of capitalization on an operating profit basis. $K$ in the denominator gives capitalization on a balance sheet basis, being directly affected by any investment or withdrawal.

A momentary rate of the rate of return on capital of Eq. (3) naturally is not representative for an entire forestry system – an expected value of the return rate is needed for management considerations. As time proceeds linearly, the expected value of the return rate on capital within a rotation can be written as [4, 5]

$$\langle r(t) \rangle = \frac{\int_b^{b+\tau} \frac{d\kappa}{dt}(t)p(t)dt}{\int_b^{b+\tau} K(t)p(t)dt} = \frac{\int_b^{b+\tau} r(t)K(t)p(t)dt}{\int_b^{b+\tau} K(t)p(t)dt} \qquad (4),$$

Where the numerator refers to the expected value of the profit rate, and the denominator to that of capitalization; $t$ is time, $p(t)$ is probability density of time within the rotation, $b$ is arbitrary starting time of the integration, and $\tau$ is rotation age. Along with a periodic boundary condition, Eq. (4) does not depend on the starting point of the integration. Value being produced in a growth process, and only changes form of appearance in any harvesting event, any computation is conducted on accrual basis.

One computational problem relates to variable timber prices. Purely random variation of prices does not contribute to the expected values of profit rates and capitalizations. Instead, an autoregressive price series evolves in time. Consequently, monetary traits no longer obey any periodic boundary condition. Physical traits may or may not obey a periodic boundary condition, but the eventual periods possibly differ from the case of stationary prices.

Within a periodic system, expected values of observables do not depend on the point of initiation of the integration of observables [Eq. (4)]. Along with the breakdown of the periodic boundary condition, the neutrality to the initial observation point obviously is lost.

Within a first-order autoregressive price evolution [Susaeta 2025, Reeves and Haight 2000], defined as

$$u(t) = u_0 + r * u(t-1) \qquad (5),$$

any price evolves as a geometric series, or

$$u(t) = u_0 \left[ 1 + \int_{t_0}^{t} z\, r^{z\,(q-t_0)} dq \right] = u_0 \left[ 1 + \frac{r^{z\,(t-t_0)} - 1}{\ln\ r} \right] \qquad (6).$$

where is $z$ a time scale factor.

Let us then discuss a tentative period of duration $\tau$, from time instant $b$ to time instant $b+\tau$. The periodic system having a continuous distribution of stages, any single stage (or phase) reappears at the instant of one full period after the previous appearance. The initial phase of the system at time instant $b$ is a priori unknown. Any phase appears once within the period, but the timing is unknown. As time is running linearly, the appearance time probability of any event is evenly distributed within the period. Correspondingly, the expected value of the current price level at system phase $s$ is

$$\langle u(s)\rangle = \int_b^{b+\tau} u(t)\ p(t)\ dt = \int_b^{b+\tau} u_0\left[1+\frac{r^{z\ (t-t_0)}-1}{\ln\ r}\right]\ p(t)\ dt = \frac{u_0}{\tau}\Big[_b^{b+\tau}\left[t\left(1-\frac{1}{\ln\ r}\right)+\frac{r^{z\ (t-t_0)}}{z\ (\ln\ r)^2}\right]$$
$$=\frac{u_0}{\tau}\left[\tau\left(1-\frac{1}{\ln\ r}\right)+\frac{r^{z\ (b-t_0)}\left(r^{z\ \tau}-1\right)}{z\ (\ln\ r)^2}\right]=u_0\left[1-\frac{1}{\ln\ r}+\frac{r^{z\ (b-t_0)}\left(r^{z\ \tau}-1\right)}{z\ \tau(\ln\ r)^2}\right] \quad (7).$$

where $p(t)=1/\tau$ is the probability density of time within the period.

The expected value of the price level at system phase *s* above contains the initial pricing $u_0$. This will naturally differ between different goods and services. This is, however, the only difference between the expected values of the prices of different monetary events. The reason is that the timing of any event is similarly evenly distributed within the period between *b* and $b+\tau$ .

## 4. Results

### *41. Results in the absence of thinnings*

Figure 3 shows the expected value of the operating profit rate from timber sales, corresponding to the numerator of Eq. (4), as a function of rotation age. It is found that the maximal expected value of the profit rate is gained with rotation of 71 years for the higher productivity, and 91 years for the lower productivity. At rotation ages below six and eleven years for the two stand productivity values, respectively, the expected values are negative since the timber sales revenues do not cover the establishment expenses.

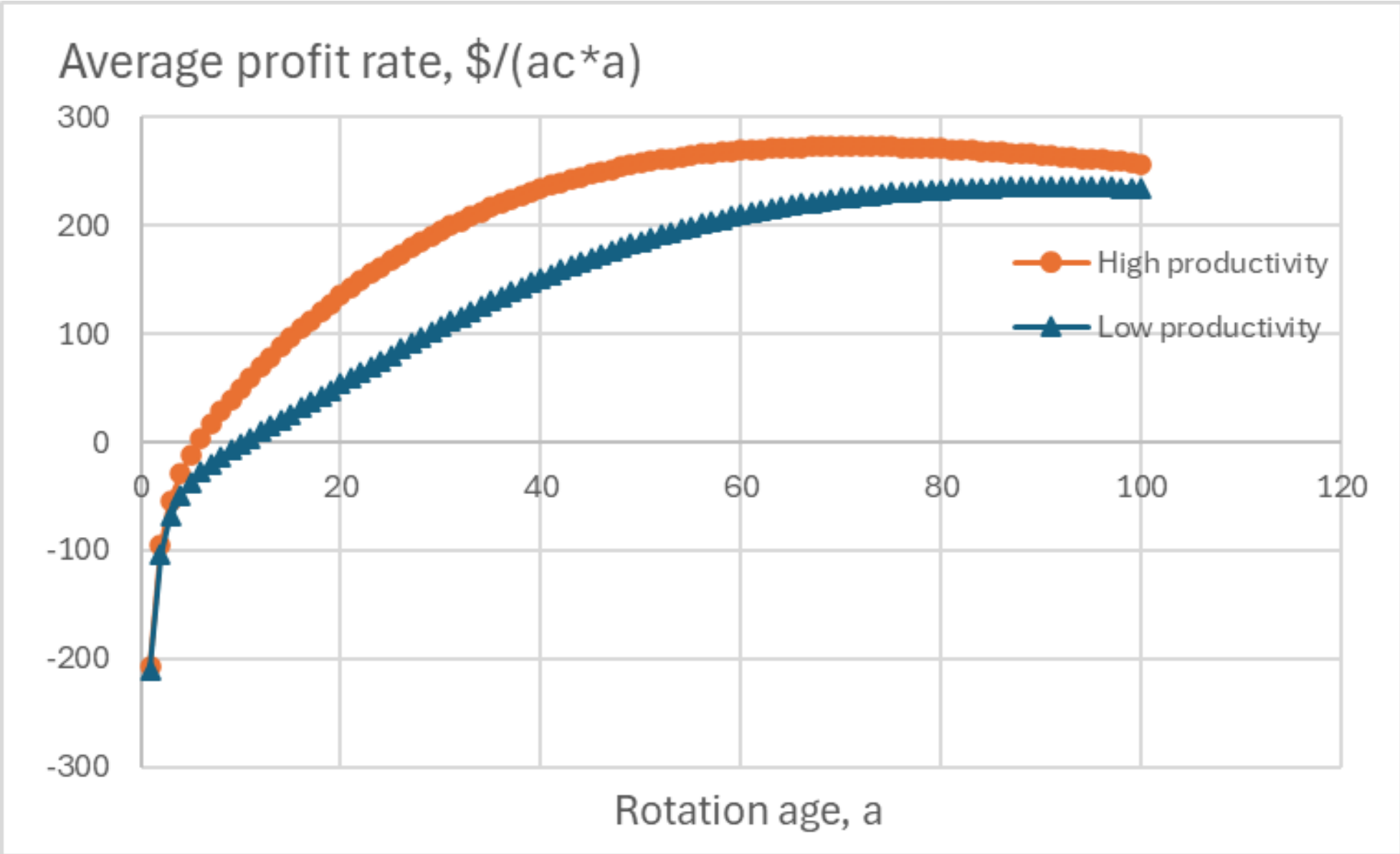

Figure 2. Expected value of the operating profit rate from timber sales as a function of rotation age.

Figure 3 shows the expected value of capitalization, corresponding to the denominator of Eq. (4), as a function of rotation age. It is worth noting that at any stand age, the capitalization on the high-productivity stand is greater, but the relative difference between stands of different productivity becomes reduced with stand age.

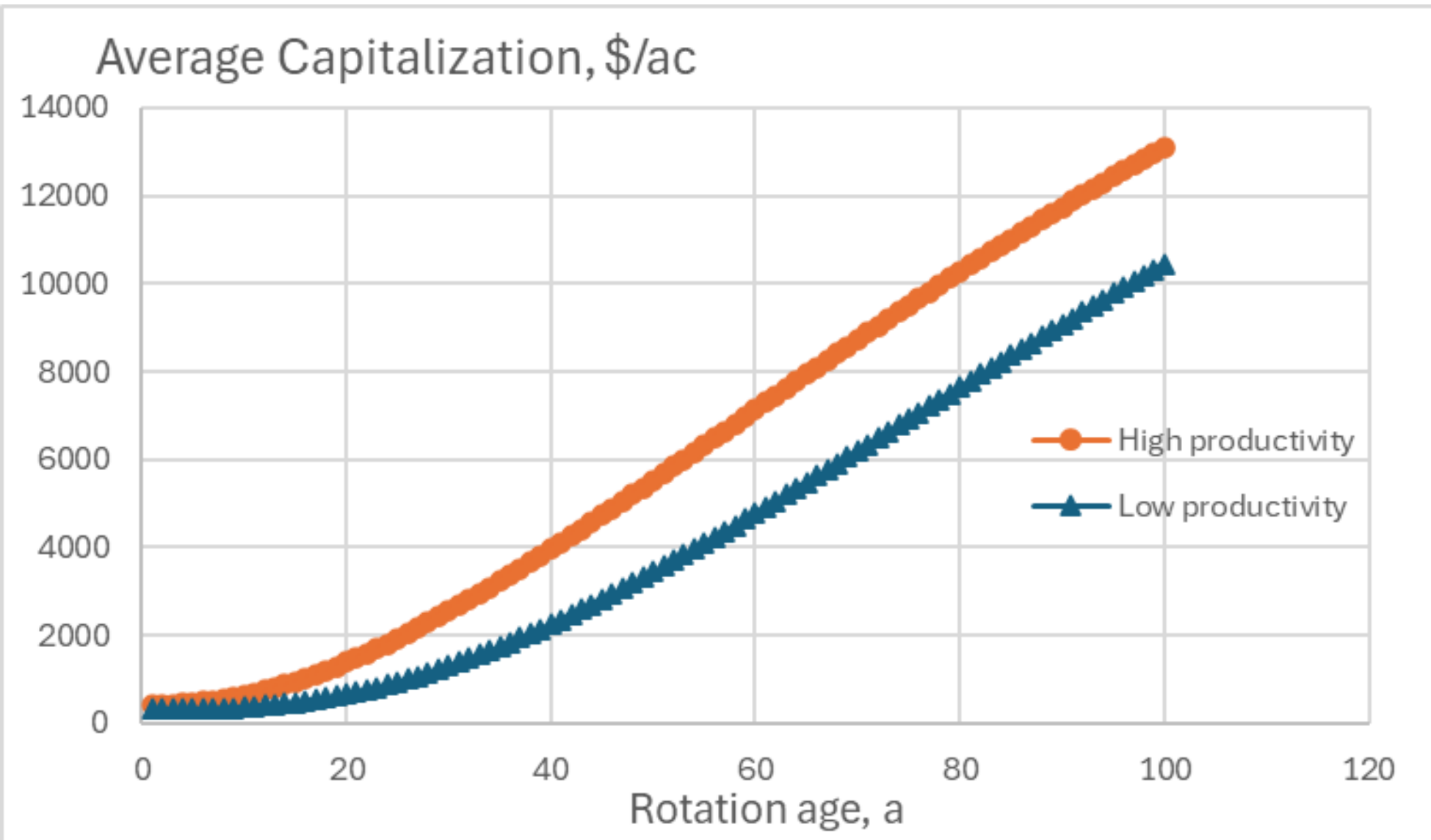

Figure 3. Expected value of capitalization as a function of rotation age.

Figure 4 shows the expected value of the return rate on capital, corresponding to Eq. (4), as a function of rotation age. It is found that the maximal expected value of the return rate is gained with rotation of 16 years for the higher productivity, and 24 years for the lower productivity. Remarkably, at long rotations, the rate of return on sites of high productivity is lower is lower than on sites of low productivity.

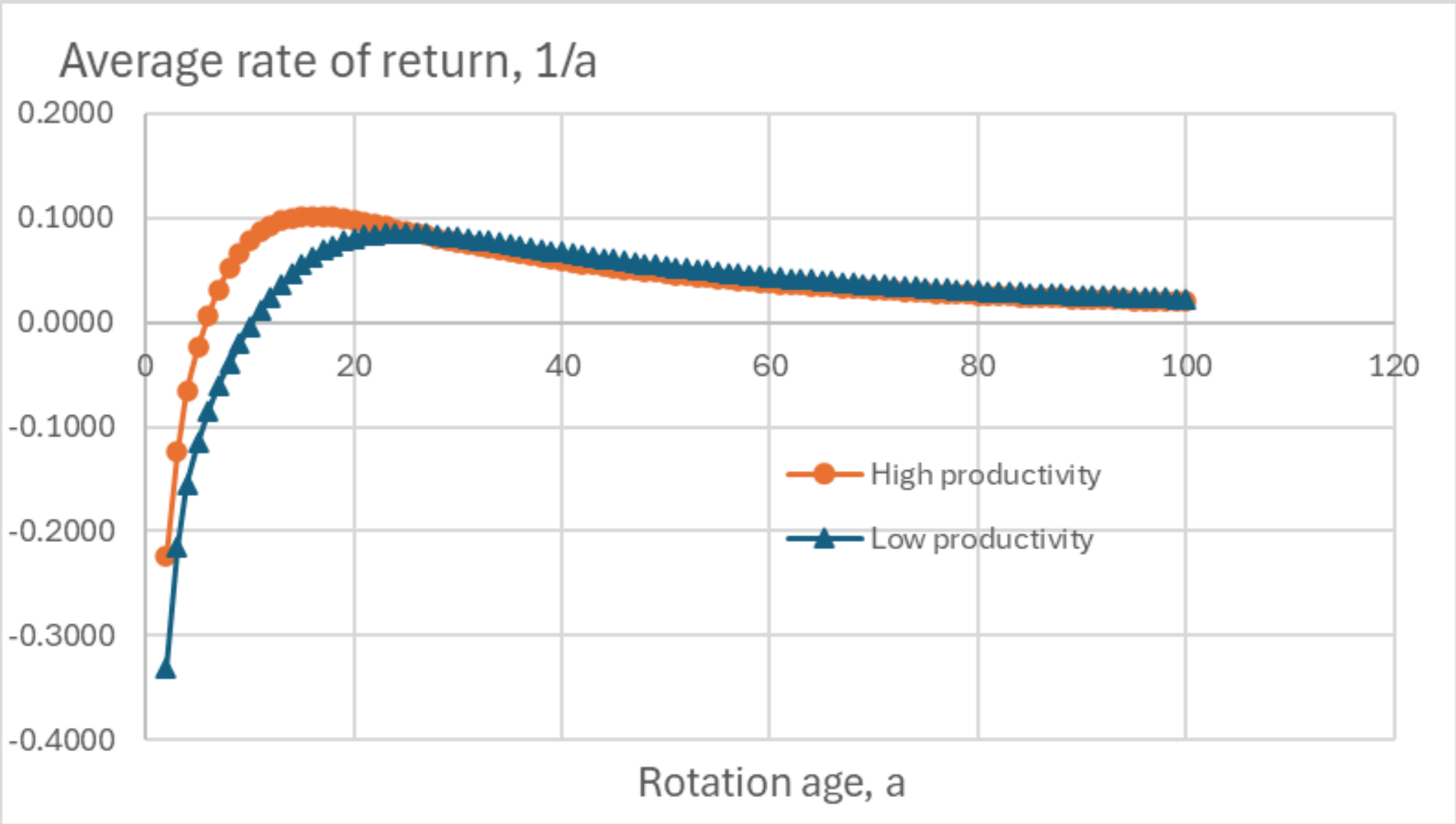

Figure 4. Expected value of the return rate on capital as a function of rotation age.

*42. Results in the presence of thinnings*

Thinnings are not feasible with values of the thinning benefit variable ("*adjustment factor*") used by Susaeta [2025]. The parameter $\delta$ in Eq. 2 must be in the order of 0.35 to make any thinning feasible. This would mean at least doubling the ("*adjustment factor*") from the value used by Susaeta [2025]. Even with greater values of the parameter $\delta$ , thinnnigs make a minor contribution to the return, and are feasible only if applied shortly before clearcutting. The reason for this adjacency requirement is that any thinning reduces volumetric yield permanently, since the correction factors in Eq. (2) do not evolve with time. Figure 5 shows the thinning effect on the rate of return on capital with $\delta = 0.9$ . In such a case a thinning benefit is visible, but the required value of the "*adjustment factor*" indeed is exaggerated.

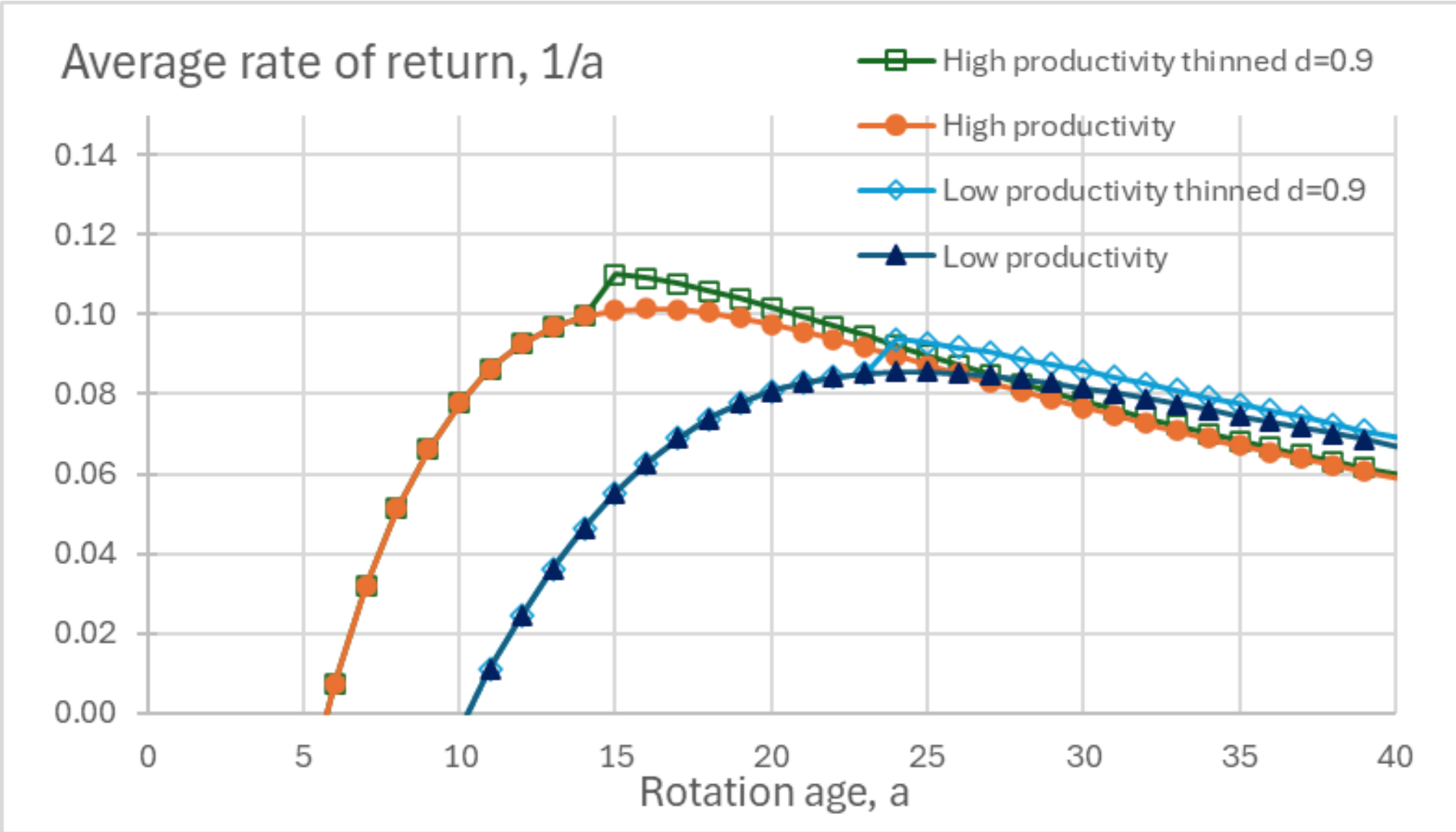

Fig. 5. Rate of return on capital with and without thinnings, as a function of rotation age. Thinnings are evaluated with an exaggerated adjustment factor $\delta$ of 0.9 in Eq. (2).

*43. Results with evolving prices and expenses*

As prices and expenses evolve according to Eq. (6), the expected values of the profit rate and the capitalization in Eq. (4) evolve. Such evolution can be investigated by changing the initiation time of the period under discussion from $b$ to $b^*$. Within a period from time instant $b^*$ to time instant $b^*+\tau$, the expected value of the profit rate is

$$\int_{b^*}^{b^*+\tau} \frac{d\kappa}{dt}(t)p(t)dt = \frac{1-\frac{1}{\ln r}+\frac{r^{z(b^*-t_0)}\left(r^{z\tau}-1\right)}{z\tau(\ln r)^2}}{1-\frac{1}{\ln r}+\frac{r^{z(b-t_0)}\left(r^{z\tau}-1\right)}{z\tau(\ln r)^2}} \int_{b}^{b+\tau} \frac{d\kappa}{dt}(t)p(t)dt \qquad (8),$$

where Eq. (7) has been applied. The expected value of the capitalization in Eq. (4) becomes

$$\int_{b^*}^{b^*+\tau} K(t)p(t)dt = \frac{1-\frac{1}{\ln r}+\frac{r^{z(b^*-t_0)}\left(r^{z\tau}-1\right)}{z\tau(\ln r)^2}}{1-\frac{1}{\ln r}+\frac{r^{z(b-t_0)}\left(r^{z\tau}-1\right)}{z\tau(\ln r)^2}} \int_{b}^{b+\tau} K(t)p(t)dt \qquad (9).$$

Equations (8) and (9) show that both profit rate and capitalization evolve with evolving prices. However, substituting both (8) and (9) into Eq. (4), the prefactors cancel. Then, the expected value of the return rate of capital does not evolve with evolving prices and expenses. In other words, the time derivative of the dimensionless quantity is retained. Consequently, the financially optimal rotation times are not affected by the evolution of prices and expenses.

## 5. Discussion

It is possible to introduce a thinning response model where the correction factors evolve in time. An example has been postulated in [Kao and Brodie 1980]. We hereby experimented with a function

$$V_a(t+1) = V(t+1) * \left\{ 1 - \frac{h(t)}{V(t)} * \exp(-d * \Delta t) \right\} \qquad (10),$$

where $d$ is decay rate of the thinning effect, and $\Delta t$ is time elapsed after thinning. Depending on the decay rate, thinnings may well become feasible, and rotation times may be greatly extended. The present dataset, however, does not allow the determination of the decay rate, and thus Eq. 10 is retained as a hypothetical possibility.

The present results differ very much from those of the recent study [Susaeta 2025] where the data was taken. Obviously, the methods are very different. Firstly, a land expectation value is computed, a method shown to induce bias in the design of periodic processes [Kärenlampi 2019, 2025a]. Secondly, a Hamiltonian is formulated for thinning dynamics [Susaeta 2025]. A Hamiltonian can be extremized if the terminal state is known. However, terminal harvesting appears to be missing from the Hamiltonian. The rotation age, as well as the terminal timber yield, possibly should depend on the thinning schedules. Finally, equality of growth and price increments with the expense rate of capital is postulated as an optimality condition [Susaeta 2025].

A remarkable finding is that in Fig. 4 is that the rate of return at long rotations is lower on sites of high productivity than on sites of low productivity. This naturally is due to the relative difference in average capitalization in Fig. 3 being greater than the relative difference in average profit rate in Fig. 2. The origin of such behavior is in the volumetric yield model given in Eq. (1), where the site fertility effect is assigned to the model parameter $c$. The site fertility effect probably would be different if the other model parameters would react to fertility.

In the present study, the evolution of expenses was treated similarly to the evolution of timber prices. This resulted in relatively simple closed-form expressions for the evolution of the profit rate and the capitalization in Eqs. (5) and (6). Then, the price level multiplier disappeared in the computation of the rate of return on capital. Yin and Newman [1995] arrived to a result like the present study, regarding to the invariance of rotation times to similarly evolving prices and expenses, even if they used very different methodology. The land expectation value (*LEV*) was affected by the evolution of prices, but expenses also evolving, rotation ages were not affected. It is natural that monetary quantities like profit rate, capitalization and *LEV* are affected by monetary multipliers. On the other hand, time derivatives of dimensionless quantities, like the return rate on capital, cannot depend on monetary multipliers.

A few investigations indicate that a relatively large timber prices reduce optimal rotation length, whereas a relatively larger expenses extend rotations [Newman et al. 1985, Rakotoarison and Loisel 2017]. Such results, however, are based on *LEV* computations, and consequently possibly include some bias [Kärenlampi 2019, 2025a]. It is, however, possible to verify such arguments in terms of a simple treatment.

For explanatory purposes, a simple stylized version of Eq. (4) is formulated as

$$\langle r \rangle \approx \frac{f\frac{dv}{dt}\tau - g}{\frac{\tau}{2}\left(f\frac{dv}{dt}\tau + g\right)} \qquad (11),$$

where the numerator contains the accumulated operating profit as the product of log price $f$, volumetric growth rate $\frac{dv}{dt}$, rotation time $\tau$, and accumulated expenses $g$ deducted. The denominator contains a linearized approximation of the sum of all capitalizations: terminal capitalization is halved and multiplied by rotation time. In Eq. (11), the bare land market value is omitted; it could easily be incorporated, but the solution would lose its instructive simplicity.

It is now possible to differentiate Eq. (11) with respect to the rotation time $\tau$, and find the maximum of Eq. (6) as the zero value of the derivative. The outcome for the optimal rotation time is

$$\tau_{opt} = \frac{g}{f\frac{dv}{dt}\left(\sqrt{2}-1\right)} \qquad (12).$$

Even if linearized Equation (11) is too simplified for the exact determination of a suitable rotation time, the implications in Eq. (12) are clear. The rotation time is proportional to the magnitude of expenses and inversely proportional to the net log price and volumetric growth rate.

The above generic argument was hereby verified using the growth model of Eq. (1), and the objective function (4). Indeed, larger expenses resulted in longer rotations, larger timber prices in shorter rotations, and vice versa.

## 6. Conclusion

The rotation times of Douglas fir forests are much shorter than in earlier results. Thinnings do not become feasible, provided the thinning effects on the volumetric yield function do not evolve in time. Evolving prices and expenses break the periodic boundary condition in monetary quantities: profit rates and capitalizations evolve with prices. The periodic boundary condition is, however, retained in time derivatives of dimensionless quantities, as well as in physical characteristics of rotations optimized for the rate of return. Then, optimal rotations do not depend on the evolution of prices and expenses. Relatively high timber prices shorten rotations, as relatively high expenses extend them.

## Funding

This work was partially funded by Niemi foundation. The funder had no role in study design, data collection and analysis, decision to publish, or preparation of the manuscript.

## Acknowledgements

N/A.

## Conflict of interest statement

The author declares that no competing interests exist.